\documentclass[a4paper,12pt]{article}
\usepackage[T1,T2A]{fontenc}
\usepackage{amsmath,amsfonts,amssymb,cite}
\usepackage{graphicx}

\begin{document}

\renewcommand*{\thefootnote}{\fnsymbol{footnote}}

\begin{center}
{\Large\bf %Polaron model for strong interactions in nucleons
Quark model of nucleon based on an analogy with polaron}
\end{center}
\bigskip

\begin{center}
{S.S. Afonin\(^{a,b}\)\footnote{E-mail: \texttt{s.afonin@spbu.ru}}
and A.V. Tulub%\(^{a}\)
}
\end{center}

\begin{center}
  {\small\({}^a\)Saint Petersburg State University, 7/9 Universitetskaya nab.,
  St.Petersburg, 199034, Russia}\\
  \vspace*{0.15cm}
  {\small\({}^b\)National Research Center "Kurchatov Institute": Petersburg Nuclear Physics Institute,
  mkr. Orlova roshcha 1, Gatchina, 188300, Russia}
\end{center}

\renewcommand*{\thefootnote}{\arabic{footnote}}
\setcounter{footnote}{0}

%\bigskip

\begin{abstract}
We demonstrate that the polaron theory from solid state physics can serve as an interesting analogue model for non-perturbative QCD,
at least in the description of nucleons and related low-energy physics of strong interactions.
By drawing explicit analogies between polaron physics, arising for an electron moving in an ionic crystal,
and physics of pion-nucleon interactions, certain rules for the "polaron/QCD correspondence" are proposed.
In polaron theory, the effective fermion mass as a function of the coupling constant is known both in the weak and strong coupling limits.
The conjectured "polaron/QCD correspondence" translates these results into strong interactions.
It is then shown how application of these rules leads to unexpectedly good quantitative predictions for the nucleon mass and the pion-nucleon sigma term.
The polaron approach also predicts that the quark degrees of freedom in the form of the constituent quark account for one-third of the nucleon mass,
consistent with lattice predictions.
We discuss possible physical reasons underlying the observed quantitative similarity between polaron physics and non-perturbative QCD.
\end{abstract}

%\bigskip

\section{Introduction}

The problem of strong coupling in low-energy strong interactions is the primary unresolved "technical" problem in the modern Standard Model. The existence and interaction of nucleons are governed by two main phenomena: confinement of color charge  and Spontaneous Chiral Symmetry Breaking (SCSB) in strong interactions, the analytical understanding of which has remained unclear for over 60 years. The standard perturbation theory in Quantum Chromodynamics (QCD) is inapplicable at strong coupling: At low energies, below 1~GeV, the effective degrees of freedom are not the quarks and gluons of the QCD Lagrangian but constituent (i.e., "dressed" by interactions) quarks and pseudo-Goldstone bosons --- primarily the pions.

It is well known that some ideas borrowed from solid state physics, in particular from the theory of superconductivity, have proven invaluable in constructing qualitative models of these phenomena. For instance, the Meissner effect and the phenomenon of Abrikosov flux tubes inspired the most popular model of quark and gluon confinement ("dual Meissner effect" in the QCD vacuum). The phenomenological Landau-Ginzburg theory of superconductivity, after relativistic generalization, led to the development of the Nambu--Jona-Lasinio (NJL) model~\cite{NJL}, which describes SCSB through the generation of a large constituent mass for nearly massless fermions (initially nucleons, later quarks~\cite{klev}) due to the formation of fermion condensate in the strong coupling regime. The interpretation of pions and other light pseudoscalar mesons as pseudo-Goldstone bosons of spontaneously broken chiral symmetry emerged in particle physics largely due to the NJL model.

In solid state physics, there is another example of a successful theory dealing with the strong coupling regime --- polaron theory, which describes, among other things, the dynamic generation of effective mass for an electron moving in an ionic crystal.
Historically, it became one of the first examples of a quantum theory describing the interaction of fermions with bosons. The origins of the polaron idea trace back to Landau~\cite{landau}, with the first version of the theory developed by Pekar in~\cite{pekar}, who also coined the term "polaron". The canonical polaron theory was constructed by Fr\"{o}hlich~\cite{Frohlich}, who initially aimed to describe superconductivity~\cite{devreese1}, several years before the BCS theory. In modern incarnations, the development of Fr\"{o}hlich's theory finds applications in attempts to describe high-temperature superconductivity (within bipolaron~\cite{devreese2,lakhno2} and spin polaron~\cite{chernyshev} models).

In the present work, we propose to apply polaron theory to the description of nucleons and some related aspects of low-energy strong interactions.
To our knowledge, there has been only one attempt in the literature to apply polaron theory to strong interactions --- the article by Iwao~\cite{ivao}, written on the wave of the "November Revolution" of 1974. In that work, the polaron mechanism was proposed to explain the formation of constituent quark mass, and following the results of polaron theory, effective coupling constants for the interaction of known quarks with gluon field were written down (with a numerical error, see footnote~2 below), without further conclusions from the assumption. We intend to go further and present various estimates in nucleon physics suggesting that further development of this idea may prove promising.

In modern theoretical physics, attempts to analytically understand the physics of strong coupling in QCD are often based on considering theories that are, in some (albeit remote) sense, related to QCD but partially allow analysis of the strongly coupled regime --- certain two-dimensional field models, supersymmetric field theories, and models based on the AdS/CFT correspondence from string theory (in its most phenomenological form, this approach is often called "AdS/QCD correspondence", a short review is given in~\cite{Afonin:2021cwo}). These "related" theories themselves do not describe any real physics. In this regard, the use of polaron theory for the aforementioned purpose, which by analogy can be called "polaron/QCD correspondence", could have a conceptual advantage: It is based on a theory that describes real physics and is experimentally verifiable. Moreover, calculations in this theory are largely possible for any coupling constant.
%, including (numerically) the intermediate regime between weak and strong coupling.

The polaron approach to QCD is certainly non-relativistic. However, it has long been known that non-relativistic models for QCD with light quarks can work well for reasons still unclear. Apparently, this is a question of a successful choice of model degrees of freedom, some thoughts on this topic are presented, for example, in~\cite{manohar,lucha}. It is quite plausible that "switching on" a sufficiently strong coupling in any initially weakly coupled relativistic quantum system will inevitably make it non-relativistic in terms of renormalized quantities (loosely speaking, an increase in the coupling constant is somewhat analogous to an increase in the viscosity parameter in a liquid) because such interactions will tend to quickly transform kinetic energy into internal potential energy. Furthermore, if a non-relativistic model is constructed that shows promising results, one can ask the question of how to "relativize" it, thereby suggesting a potentially promising direction for further research.

The work is organized as follows. Section 2 briefly outlines some basic results from polaron theory, focusing only on those aspects that are later applied to strong interactions in Section~3. Possible relationships between strong interactions and polaron physics are further discussed in Section~4. The concluding Section~5 summarizes the findings. A heuristic discussion of a possible semi-analytical connection between the obtained model results and QCD is given in the appendix.

\section{Some results from polaron theory}

In ionic crystals (or polar dielectrics), atoms in the lattice sites are relatively weakly bound compared to covalent bonds, so an electron or other charge carrier moving in such an environment not only locally polarizes the environment but also induces intense lattice vibrations, which in turn affect the moving electron. The resulting back reaction is effectively described as the interaction of the electron with quanta of lattice vibrations --- phonons. An electron moving together with a "cloud" of surrounding phonons is called a polaron. Due to the electron-phonon interactions, such an electron is forced to "drag" along with it the lattice deformation caused by its presence, leading to a significant increase in the effective mass of the resulting quasiparticle compared to the mass of the electron in the conduction band.

Polarons have been extensively studied (see, for example, references in the reviews~\cite{devreese1,devreese2,lakhno}). In short, polarons are divided into acoustic and optical. In the former case, neighboring atoms in the lattice vibrate in phase --- such vibrations can interact with sound waves in the medium. In the latter case, when all or some neighboring atoms vibrate out of phase, the polarons are optical. They, in turn, can be of two types: Transverse optical polarons (vibrations of charges in the lattice sites are asymmetric\footnote{Corresponding images can easily be found online using relevant keywords.}, leading to a varying dipole moment that can interact with a transverse electromagnetic wave --- the resulting quasiparticles are called phonon polaritons) and Longitudinal Optical Polarons (LOP) (the charge distribution during vibrations remains symmetric, no dipole moment is formed). The interaction of the electron with phonons is strongest in the case of LOP --- in typical polar crystals, the corresponding dimensionless effective coupling constant is 3--6~\cite{lee,tulub}, which is much larger than the electron-photon interaction constant $1\!/\!137$. As a result of this strong interaction at large distances (compared to the distance between lattice nodes), the electron acquires a large effective mass $m^*$, which can be tens of times greater than the mass of an electron in the conduction band. Thus, LOP is a very interesting example of an emergent phenomenon where the effective interaction arising in the structure at "large" distances can be in the strong coupling regime, while the underlying fundamental interactions at "small" distances are in the weak coupling regime.

An important feature of LOP is that the frequency $\omega_0$ of the corresponding phonon is almost independent of its wave vector $k$, and it can be considered constant to a good approximation. This simplifies the construction of LOP theory. Another, more important simplification arises from the observation that the effective size of LOP (the size of the polaron-induced lattice deformation) is generally much larger than the lattice constant. This allows replacing the lattice with a continuous polarizable medium through which the electron moves. Under these two assumptions, and neglecting spin and relativistic effects, Pekar proposed the first theory for such a "large" LOP in~\cite{pekar}. Later, Fr\"{o}hlich justified the now-standard Hamiltonian defining this theory~\cite{Frohlich},
\begin{equation}
\label{1}
H=-\frac{\hbar^2}{2m}\triangle_r+\sum_k\hbar\omega_0 a_k^\dag a_k+\sum_k \left(V_k a_ke^{ikr}+h.c.\right).
\end{equation}
Here, $r$ is the coordinate operator of an electron with mass $m$ in the conduction band of an ionic crystal or polar dielectric (i.e., $-\hbar^2\triangle_r=p^2$, where $p$ is the momentum operator of the electron), $a_k^\dag$ and $a_k$ are the creation and annihilation operators of LOP with energy $\hbar\omega_0$ and three-dimensional wave vector $k$,  $[a_k,a_{k'}^\dag]=\delta_{kk'}$, and $V_k$ is a specific function of $k$. In ionic crystals~\cite{devreese1},
\begin{equation}
\label{1a}
V_k = -i\frac{\hbar\omega_0}{|k|} \left(\frac{4\pi\alpha}{V}\right)^{1/2}\left(\frac{\hbar}{2m\omega_0}\right)^{1/4},
\end{equation}
\begin{equation}
\label{1b}
\alpha = \frac{e^2}{\hbar c} \left(\frac{mc^2}{2\hbar\omega_0}\right)^{1/2}\left(\frac{1}{\varepsilon_\infty}-\frac{1}{\varepsilon_0}\right),
\end{equation}
where $\alpha$ is the electron-phonon coupling constant, $\varepsilon_\infty$ and $\varepsilon_0$ are the high-frequency and static dielectric constants, and $V$ is the system volume. In practice, $\varepsilon_\infty$ (determined only by the polarization of the electronic shells of ions, as the position of ions in the lattice hardly changes in a high-frequency electric field) is several times smaller than $\varepsilon_0$, so $\alpha>0$.

The problem defined by the Hamiltonian~\eqref{1} is solved using the variational method. There are many works on this topic in the literature (see, for example, the reviews~\cite{devreese1,devreese2,lakhno}), which can be roughly divided into two main approaches. In the first approach, two canonical transformations are made: The first (Heisenberg transformation) formally removes the dependence on the electron coordinates, and the second, proposed by Lee, Low, and Pines~\cite{lee}, effectively separates the Hamiltonian into classical and quantum parts, which can be interpreted as "dressing" the electron with a field of virtual phonons describing polarization~\cite{devreese1}. The second transformation introduces trial functions, after which the resulting variational problem for the transformed Hamiltonian $\tilde{H}$ is solved. The second approach to solving the polaron problem was introduced by Feynman in~\cite{feynman}. In this approach, the problem is first reformulated in Lagrangian terms, a path integral is introduced, and then integration over the phonon fields is performed, resulting in a quantum-mechanical path integral over the electron's trajectories $X(t)$. This technique was invented by Feynman several years earlier in one of his works on the formulation of quantum electrodynamics. The final action has the form
\begin{equation}
\label{2}
S = \frac12\int \!\left(\frac{dX}{dt}\right)^2\!dt + \frac{i\alpha}{2^{3/2}}\int\!\!\int\!\frac{e^{-i|t-s|}}{|X_t-X_s|}\,dtds.
\end{equation}
The variational approach is then applied to the action~\eqref{2}. The results of the calculations turned out to be close to those of the first approach.

The first solution for the theory defined by the Hamiltonian~\eqref{1} in the weak coupling limit was proposed by Fr\"{o}hlich, and in the strong coupling limit by Pekar. Pekar used the adiabatic approximation (the electron adiabatically follows the motion of the atoms), within which the wave function $\Psi$ of the "electron + phonon field" system factorizes as
\begin{equation}
\label{3}
\Psi(r,\{q_i\})=\psi(r)\Phi(\{q_i\}),
\end{equation}
where $\psi(r)$ is the electron wave function, depending only on the electron coordinates, and $\Phi$ is the wave function of the phonon field, depending only on the field coordinates $\{q_i\}$. The factorization~\eqref{3} in the strong coupling limit has become a common assumption in polaron theory. Note that the ansatz~\eqref{3} does not preserve the translational symmetry of the original theory~\eqref{1}, thus predicting the localization of the polaron wave function. The same property holds for Feynman's theory~\eqref{2}~\cite{feynman}, which provides a solution in both limits, smoothly connecting them. An ansatz that does not violate translational symmetry was first proposed in~\cite{tulub}, where the corresponding solution in both limits was also obtained. This ansatz is based on the diagonalization of the quantum part of the Hamiltonian $\tilde{H}$, proposed earlier in~\cite{tulub3}.

The concepts of strong and weak coupling are defined by the ratio of the ground state energy $E_0$ of the electron in the potential well created by the polaron (i.e., the energy of the lattice's electric polarization) to the LOP energy $\hbar\omega_0$, where $\omega_0$ is its frequency, independent of the wave vector. The effective mass of the polaron $m^*$ is determined from the kinetic energy $m^*\!v^2\!/2$ of a slow electron moving in the crystal with an average group velocity $v$. The value of $m^*$ can be directly measured in experiments using cyclotron resonance. In the weak and strong coupling limits, polaron theory~\eqref{1} leads to the following predictions for $E_0$ and $m^*$~\cite{devreese1},
\begin{eqnarray}
\label{4}
   \frac{E_0}{\hbar\omega_0}&=&-\alpha-0.01\underline{6}\alpha^2-\dots \quad (\alpha\rightarrow0),
   \\
   \label{5}
   \frac{E_0}{\hbar\omega_0}&=&-0.10\underline{6}\alpha^2-\dots \qquad\quad (\alpha\gg1),
\end{eqnarray}
\begin{eqnarray}
\label{6}
   \frac{m^*}{m}&=&1+\frac{\alpha}{6}+0.02\underline{4}\alpha^2+\dots \quad (\alpha\rightarrow0),
   \\
   \label{7}
   \frac{m^*}{m}&=&0.02\underline{0}\alpha^4+\dots \qquad\qquad\quad (\alpha\gg1),
\end{eqnarray}
where the underlined numbers vary slightly among different authors.

The translationally invariant ansatz of~\cite{tulub} initially predicted a bit larger value for the ground state energy in the strong coupling regime, $E_0=-0.105\alpha^2\hbar\omega_0$, but almost half a century later it was found~\cite{tulub2} that the variational calculation in~\cite{tulub} used an oversimplified trial function (with one parameter), violating Pekar's virial relation~\cite{pekar} which must hold in the polaron problem. After refining the ansatz, the predicted ground state energy turned out to be lower than in~\eqref{5},
\begin{equation}
\label{8}
\frac{E_0}{\hbar\omega_0}=-0.1257520\alpha^2 \qquad\quad (\alpha\gg1).
\end{equation}
This result clearly indicates that a strongly coupled polaron without clear spatial localization is energetically more favorable. A rigorous justification for the absence of such a localization of the LOP wave function at any coupling constant was given in~\cite{gerlach}.

\section{Application to strong interactions}

What can polaron physics have to do with strong interactions? Already Feynman expressed in~\cite{feynman} the desire to apply his polaron theory~\eqref{2}, after a proper relativization, to the construction of a theory of mesons.

Note that the Hamiltonian~\eqref{1} has a rather general form. Historically, the construction of first theories for strong interactions began in the 1940s with the analysis of pion-nucleon interaction Hamiltonians, which look very similar to~\eqref{1}~\cite{pauli}. In them, the role of phonons was played by pions, while the interaction between pions was neglected, as in the case of phonons in the Hamiltonian~\eqref{1}. The developments from those years are collected in the review~\cite{wick}. Briefly, the Hamiltonian of the pion-nucleon system had the form
\begin{equation}
\label{pn1}
H=H_0+\sum_{k,j} \left(V_k^j a_k^j e^{ikr}+h.c.\right),
\end{equation}
where $H_0$ is the Hamiltonian of free pions and nucleons, the latter usually considered infinitely heavy. The linearity of~\eqref{pn1} with respect to $a_k^j$, as usual, is equivalent to the assumption that the emission and absorption of quanta occur sequentially (here $j$ is the isospin index). The transition matrix element $V_k$, according to the notation of the review~\cite{wick}, is equal to
\begin{equation}
\label{pn2}
V_k^j = iN\! \left(4\pi\right)^{\frac12}\!\left(f\!/\!\mu\right)\tau^j \vec{k}\vec{\sigma}v(k)\left(2\omega_k\right)^{-\frac12}.
\end{equation}
Here, $N$ is a normalization factor, $f$ is a dimensionless coupling constant, $\mu$ is a dimensional parameter, and the function $v(k)$ is subject to the condition $v(k)\rightarrow0$ as $k\rightarrow\infty$, which removes divergences from the emission and absorption of virtual quanta with large momentum. The expression~\eqref{pn2} takes into account the conservation of spin and isospin during the interaction (through the matrices $\vec{\sigma}$ and $\tau^j$) and the fact that pions are emitted and absorbed only in the $p$-wave due to the conservation of spatial parity by strong interactions. Based on the Hamiltonian~\eqref{pn1}, the concept of a "dressed particle" (introduced by Yu.V. Novozhilov~\cite{nov1,nov2}) emerged: An elementary particle together with its surrounding cloud of virtual particles. Later, the most important example of a dressed particle became the constituent quark.

In the review~\cite{wick}, the origin of~\eqref{pn1} from field theory is discussed in detail. In modern terms, the corresponding interaction Hamiltonian density can be qualitatively reformulated as
\begin{equation}
\label{pn3}
\mathcal{H}_\text{int}=\left(f\!/\!\mu\right) \bar{\Psi}\gamma^\nu\gamma^5\tau^j\Psi\partial_\nu\pi^j.
\end{equation}
The fermions in~\eqref{pn3} interact with the derivative of the pion field as it must be for the Goldstone boson.
Next, it is necessary to expand the field $\pi^j$ in a Fourier series, consider the static approximation, $\partial_t\pi^j=0$, and account for the finite size of the nucleon by effectively averaging over it --- the dimensional parameter $\mu$ is specifically related to this size. It would be interesting to analyze in more detail the connection between the Hamiltonians~\eqref{1} and~\eqref{pn1}, in particular, one can hypothesize different mechanisms leading to the $1/|k|$ scaling for $V_k^j$ in~\eqref{pn1} to conform it with~\eqref{1a}, but this would lead us too far from the main line. In this work, we will limit ourselves to the general assumption that such a connection exists and discuss its possible implications.

To our knowledge, the only attempt to apply polaron theory~\eqref{1} to strong interactions was undertaken (without any justification) in the work of Iwao~\cite{ivao}, published in 1976. In that work, it was suggested that the role of the lattice node is played by the center of mass of the hadronic "bag", the polarizable medium consists of sea quarks, and the electron-phonon interaction is replaced by the interaction of gluons with valence quarks. Since then, the understanding of strong interactions has matured significantly.

Let us replace the nucleon fields in the Hamiltonian of type~\eqref{pn3} with the quark fields inside the nucleon. The size of current quarks in the Standard Model is known to be less than $10^{-3}$ fm, while the size of hadrons is about $1$ fm. In this sense, the nucleon appears as a large object for the current quark, hence, at sufficiently small time scales, its motion within the nucleon may indeed resemble motion in a polarizable (with respect to color charge) medium. Gluons are perturbative objects and it is more reasonable to consider gluon interactions as the mechanism that sustains the system --- in the same sense that electromagnetic interactions are responsible for the existence of an ionic crystal lattice. In the latter, the electron-phonon interaction at large distances is introduced as a new effective interaction, whose connection to photons is highly complex and is considered to be encoded in the parameters of the model Hamiltonian describing this interaction. In QCD, strong coupling also arises at large distances, where the interaction is primarily reduced to pion exchanges.

Given this fact, as well as the aforementioned possible similarity between the Hamiltonians~\eqref{1} and~\eqref{pn1}, it seems reasonable to consider the interaction of pions with valence quarks as analogous to the electron-phonon interaction. Moreover, like phonons, pions do not carry spin. Also like phonons, pions represent collective excitations of a "medium", the role of which is played by the QCD vacuum (more accurately, pions are often interpreted in this way because of their pseudo-Goldstone nature in strong interactions). This comparison leads to a natural analogy for the universal frequency $\omega_0$ of the longitudinal optical polaron, independent of the wave vector --- this is the pion mass $m_\pi$ (in units $\hbar=c=1$) since $m_\pi$ is the only natural energy scale associated with pion exchanges.

It is worth emphasizing this point once more. The classical polaron theory~\eqref{1} is based on the assumption that periodic crystal can be replaced by a continuum medium in which an electron propagates and interacts with phonons (no explicit photons). The fluctuating gluon field inside a hadron can also be modeled as a "continuum medium" in which a quark propagates and interacts with pions (no explicit gluons). The resulting effective Hamiltonians of these two models may look very similar or even identical in some limiting case.

Non-perturbative strong interactions are believed to effectively convert the current valence quarks in the nucleon into constituent quarks with a large effective mass and size comparable to the mass and size of observed nucleons. At low energies, nucleons interact as objects surrounded by a cloud of virtual pions. Moreover, an explicit introduction of the meson cloud is necessary for a correct description of the electromagnetic form factors. The non-perturbative structure of nucleons in this situation could be modeled as a superposition of virtual pions with different wavelengths surrounding the valence quarks. This has some analogy in solid state physics: The internal structure of the polaron can be represented as a potential well formed from a set of optical phonons with different wavelengths due to strong electron-phonon interactions.

As was discovered at the dawn of the study of low-energy strong interactions, the phenomenological constant of the pion-nucleon interaction $\alpha_{\pi\! N\!N}$ is of the order of  $15$~\cite{Halzen}. Results from later experiments led to values lying approximately in the interval $\alpha_{\pi\! N\!N}=14\pm1$~\cite{ericson}, with the most modern value being $\alpha_{\pi\! N\!N}=\frac{g^2_{\pi\! N\!N}}{4\pi}\approx13.9$, where $g_{\pi\! N\!N}\approx13.2$~\cite{epelbaum}. This value is much larger than the coupling constant $\alpha_s$ in QCD. The constant $\alpha_{\pi\! N\!N}$, by definition, enters into the relation for the pion-nucleon scattering cross-section in the same way as the fine-structure constant enters into the Rutherford scattering cross-section of a photon on an electron.
The coupling $\alpha_{\pi qq}$ between constituent quarks and pions should be of the same order as $\alpha_{\pi\! N\!N}$ but the exact ratio depends on the model. For instance, in the
$SU(4)$ limit of spin-flavor symmetry, the constants of the pion-nucleon and pion-quark interactions are related by $g_{\pi qq}=\frac{3}{5}g_{\pi\! N\!N}$~\cite{riska}.

An electron placed in an ionic lattice becomes not only "dressed" in a polaron, but also acquires a certain binding energy of its ground state. Similarly, a quark "placed" in the QCD vacuum becomes not only "dressed" in a constituent quark, but also acquires a certain binding energy of its ground state. We will identify the absolute value of this energy with the mass of the nucleon. In this simplified picture, the other two valence quarks are interpreted as a part of the surrounding "medium", strong interaction with which neutralizes the color charge of original quark. In the simplest case, the underlying mechanism is described as the formation of a spin-singlet diquark $[u,d]$ with opposite color charge~\cite{wil}.

Summarizing the above assumptions, we can formulate the rules of "polaron/QCD correspondence", which are presented in Table~1.

\begin{table}
\centering
\caption{The rules of conjectured "polaron/QCD correspondence".}
\vspace{0.5cm}
\begin{tabular}{|c|c|}
\hline
\bf Longitudinal Optical Polarons & \bf Strong Interactions \\
\hline
\hline
Photons & Gluons \\
\hline
Phonons & Pions \\
\hline
Electrons & Quarks \\
\hline
Polaron & Constituent quark \\
\hline
Electron mass $m$ in conduction band  & Current quark mass $m_q$ in QCD  \\
\hline
Effective electron mass $m^*$ & Constituent quark mass $m_q^*$ \\
\hline
Ionic lattice & Virtual gluons and quark-antiquark pairs,\\
& neighboring valence quarks \\
\hline
Absolute value of electron ground & Ground state energy of dressed quark \\
state energy $E_0$ in ionic lattice & in QCD vacuum = nucleon mass $M_N$ \\
\hline
Electron-phonon coupling $\alpha$ & Quark-pion coupling $\alpha_\text{eff}$ \\
\hline
Phonon energy $\hbar\omega_0$ &  $m_\pi c^2$ \\
\hline
\end{tabular}
\end{table}

Accepting these assumptions, let us examine the numerical estimates provided by formulas~\eqref{4}--\eqref{8} in the context of strong interactions.

First, consider the weak coupling limit. This limit in QCD corresponds to energy scales at which perturbation theory is applicable --- up to values of the QCD coupling constant $\alpha_s$
in the range 0.3--0.4. The largest values of $\alpha_s$ measured experimentally lie in this interval~\cite{pdg}. In the non-relativistic limit, the coupling $\alpha_s$ with very similar
phenomenological values appears directly in the Cornell potential describing the interquark forces~\cite{bali},
\begin{equation}
V(r)=-\frac{\kappa}{r}+\sigma r+C,
\end{equation}
where
\begin{equation}
\kappa =C_F\,\alpha_s,\qquad C_F= \left\{
\begin{array}{cc}
4/3&\text{for mesons}\\
2/3&\text{for baryons}.
\end{array}
\right.
\label{cas}
\end{equation}

The mass scale of light baryons roughly corresponds to the aforementioned range of strong coupling and we identify $\alpha_s$ with $\alpha_\text{eff}$ in the limit of weak coupling.
For typical bare masses of light quarks, relation~\eqref{6} does not lead to any noticeable generation of effective dynamical mass.

For $\hbar\omega_0 \equiv m_\pi=140$~MeV~\cite{pdg}, relation~\eqref{4} yields
\begin{equation}
\label{8b}
-E_0\approx\text{40--60~MeV}.
\end{equation}
By assumption, the absolute value of $E_0$ should give the nucleon mass. Then a small value~\eqref{8b} is predicted in the weak coupling limit.

We can propose the following physical interpretation. If the nucleon mass is defined via the QCD energy-momentum tensor (see~\eqref{a2b} in the appendix), then the contribution to the mass can be conditionally divided into two parts --- the first reflects the contribution from non-perturbative gluon interactions, and the second
(see~\eqref{a10}) gives the so-called pion-nucleon sigma term $\sigma_{\pi N}$~\cite{al}, which describes the contribution from light quarks arising due to SCSB. In the strong coupling regime, the first contribution absolutely dominates. But if we theoretically take the weak coupling limit, that contribution should become insignificant. The contribution from the sigma term should probably not change much, since it does not explicitly depend on the coupling constant, describing some analogue of the Higgs mechanism in strong interactions. Modern estimates of the sigma term lie in the range 40--60~MeV~\cite{al}, exactly this range is reproduced in~\eqref{8b}! In other words,
within this interpretation, the asymptotic result~\eqref{4} of polaron model yields a very simple relation for the pion-nucleon sigma term,
\begin{equation}
\label{8c}
\sigma_{\pi N}\simeq\alpha_s m_\pi,
\end{equation}
where $\alpha_s$ should be taken near the scale of nucleon mass.

Next, let us consider the more interesting case of strong coupling. The low-energy interactions of the lightest $u$ and $d$ quarks, including the formation of stable nucleons, are almost entirely determined by the strong coupling regime in QCD. According to Gell-Mann--Oakes--Renner (GOR) relation~\cite{Gell-Mann:1968hlm},
\begin{equation}
\label{GOR}
m_\pi^2f_\pi^2=-(m_u+m_d)\langle \bar{q}q\rangle,
\end{equation}
where $\langle \bar{q}q\rangle\equiv\langle \bar{u}u\rangle=\langle \bar{d}d\rangle$ and $f_\pi=92.2$~MeV~\cite{pdg}, the sum of the masses of the lightest quarks at the pion mass scale is $m_u+m_d\approx11$~MeV for the standard phenomenological value of the quark condensate, $\langle \bar{q}q\rangle\approx-(250~\text{MeV})^3$ (more precisely, the chiral symmetry arguments suggest $m_u\approx4$~MeV and $m_d\approx7$~MeV~\cite{gasser,gasser2}, then
$\langle \bar{q}q\rangle=-(247.4~\text{MeV})^3$ for $m_\pi=140$~MeV). Thus, it seems natural to set $m_q=\overline{m}_{u,d}\approx5.5$ MeV. A difference of almost two orders of magnitude compared to the constituent mass gives a
qualitative idea of the strength of low-energy strong interactions.

A quick glance at relations~\eqref{5} and~\eqref{7} shows that the scaling $E_0\sim\sqrt{m^*}$ takes place at strong coupling. This scaling is a non-trivial prediction of the polaron model.
More precisely, ignoring the underlined numbers, one can write
\begin{equation}
E_0^2\sim \frac12\frac{m^*}{m}(\hbar\omega_0)^2.
\end{equation}
Now making use of the correspondence from Table~1 and GOR relation~\eqref{GOR} we obtain a remarkable prediction
\begin{equation}
\label{main}
M_N^2\sim\frac12\frac{m_q^*}{m_q}m_\pi^2=-\frac{m_q^*\langle \bar{q}q\rangle}{f_\pi^2}.
\end{equation}
In the standard constituent quark model, the nucleon mass arises from the masses of constituents, to a first approximation, from a trivial summation $M_N\approx 3m_q^*$.
Such a summation, however, is neither relativistic nor renormalization scale invariant. An example of relativistic and renormalization scale invariant way of summation over
masses of constituents is given by GOR relation~\eqref{GOR}. In this sense, relation~\eqref{main} extends (up to a factor) GOR relation to the case of the nucleon mass.
The key difference is the replacement of the current quark mass in GOR relation by the constituent quark mass.
Under certain assumptions, the scaling $M_N^2\sim -m_q^*\langle \bar{q}q\rangle$ can be obtained from the energy-momentum tensor in QCD.
We demonstrate this (partly based on the discussions in Ref.~\cite{ahep}) in the appendix.

Polaron model also predicts a common factor in~\eqref{main}: Taking into account the numerical factors in relations~\eqref{5} and~\eqref{7}, one can write
\begin{equation}
\label{c11}
M_N^2=\frac{C^2 m_\pi^2}{0.02}\,\frac{m_q^*}{m_q}=-(10C)^2\frac{m_q^*\langle \bar{q}q\rangle}{f_\pi^2},
\end{equation}
where the factor $0.02$ from~\eqref{7} was substituted and the constant $C$ is the absolute value of numerical factor appearing in~\eqref{5} or~\eqref{8}.
Substituting $C=0.1257520$ from~\eqref{8} we get a refined prediction
\begin{equation}
\label{11d}
M_N^2=\frac{1.58\, m_\pi^2}{2}\,\frac{m_q^*}{m_q}=-\frac{1.58\,m_q^*\langle \bar{q}q\rangle}{f_\pi^2},
\end{equation}
or approximately
\begin{equation}
\label{11d2}
M_N^2\approx-\frac{3m_q^*\langle \bar{q}q\rangle}{2f_\pi^2}.
\end{equation}
In the appendix, we discuss how~\eqref{11d2} could arise in QCD.

For numerical prediction of the nucleon mass, we need to know the value of the constituent mass $m_q^*$.
This value is a model-dependent quantity, differing from $\frac{1}{3}$ of the nucleon mass (for non-strange quarks) by many tens of MeV in various models.
Many of these models are based on a certain parameterization of relativistic effects within initially non-relativistic approaches,
so it is unclear which value should we prefer in our estimates. As a reasonable guide (and the simplest possibility) let us just take $\frac{1}{3}$ of the mean mass of the proton and neutron,
\begin{equation}
\label{cons}
m_q^*=\frac13\frac{M_p+M_n}{2}\approx\frac{939}{3}=313\,\text{MeV}.
\end{equation}
Actually a very close value of constituent quark mass appears in many potential models and also typically arises in the lattice simulations and calculations based on the Schwinger--Dyson equations (see, e.g.,~\cite{Aguilar:2012rz}).
With our inputs, substituting~\eqref{cons} to~\eqref{11d} yields $M_N\approx939$~MeV --- almost precisely the mean nucleon mass!

And {\it vice versa}: If we fix the nucleon mass, a typical value of the mass of the constituent quark will be obtained.

Finally let us estimate the coupling constant $\alpha_\text{eff}$.
Substituting our $m_q$ and $m_q^*$ into~\eqref{7}, we get
\begin{equation}
\label{9}
\alpha_\text{eff}\approx\left(\frac{313}{0.02\cdot5.5}\right)^{1/4}\approx7.3.
\end{equation}
This value is approximately two times smaller than the phenomenological constant of the pion-nucleon interaction\footnote{Note that in the work of Iwao~\cite{ivao}, a similar estimate for $\alpha_\text{eff}$ in the case of $u$ and $d$ quarks is approximately two times larger than in~\eqref{9}, and thus (accidentally) close to the experimental $\alpha_{\pi\! N\!N}$. The reason for the discrepancy is rather funny. Feynman, having obtained the numerical result~\eqref{7} in his work~\cite{feynman}, seems to interpolated (apparently for the sake of "beauty") the number $0.02$ by $\left(\frac{2}{3\pi}\right)^4=\frac{16}{81\!\pi^4}$. However, it is easy to verify that $\frac{16}{81\!\pi^4}\approx0.002$. This "beautiful" version of the result was used in~\cite{ivao}, which led to an additional factor of $10^{1\!/\!4}\approx1.8$ in the estimate of $\alpha_\text{eff}$ from~\eqref{7}.} $\alpha_{\pi\! N\!N}$ and thus gives a reasonable
estimate for $\alpha_{\pi qq}$.
%The given ratio is close to the relation $\alpha_{\pi qq}=\frac{3}{5}\alpha_{\pi\! N\!N}$~\cite{riska}.

The coupling $\alpha_\text{eff}$ can also be estimated in a completely different way. Relation~\eqref{1b} predicts the following scaling~\cite{tulub},
\begin{equation}
\label{9b}
\alpha=\alpha_0\sqrt{\frac{m^*}{m}},
\end{equation}
where $\alpha_0$ is the value of the coupling constant in the case where the effective mass $m^*$ equals the mass $m$ of the electron in vacuum. Applying the "polaron/QCD correspondence" from Table~1, with our input parameters, we obtain
\begin{equation}
\label{10}
\alpha_\text{eff}\approx\sqrt{\frac{313}{5.5}}\,\alpha_s\approx7.6\,\alpha_s,
\end{equation}
where $\alpha_s$ is the coupling constant in perturbative QCD. According to perturbation theory, $\alpha_s$ grows without bound as the transferred momentum decreases. However, perturbation theory does not work for $\alpha_s\gtrsim1$. There are many arguments suggesting that, in reality, as the transferred momentum approaches $\Lambda_\text{QCD}$, $\alpha_s$ "freezes" near the value $\alpha_s^\text{\tiny (fr)}\approx1$, reaching this asymptote already at the pion mass scale (see, for example, Fig.~4.1 in the review~\cite{deur} with the normalization of $\alpha_s$ adopted there, which differs by a factor of $\pi$ from the commonly accepted one in Particle Data~\cite{pdg}, or Fig.~3.2 in the same review~\cite{deur}). This estimate of the infrared-fixed value of $\alpha_s^\text{\tiny (fr)}$ is quite rough and model-dependent; nevertheless, substituting it into~\eqref{10}, we arrive at an alternative estimate, $\alpha_\text{eff}\approx7.6$, which is quite close to the estimate~\eqref{9}.

The last but not least remark: Connecting~\eqref{9b} with~\eqref{11d}, we get an approximate relation
\begin{equation}
\label{10b}
M_N\approx\frac{\alpha_\text{eff}}{\alpha_s^\text{\tiny (fr)}}\,m_\pi,
\end{equation}
i.e., at the "frozen" strong coupling $\alpha_s^\text{\tiny (fr)}\approx1$, the value of effective coupling $\alpha_\text{eff}$ in the polaron approach determines the ratio $M_N/m_\pi$.
Moreover, relation~\eqref{10b} has then the same form as~\eqref{8c}, $M_N\approx\alpha_\text{eff}m_\pi$, this gives a simple qualitative picture for dynamical generation of the nucleon mass from the pion-nucleon sigma-term.

\section{Discussions}

When describing quantum systems by means of theoretical models, the intensity of interactions depends on the choice of physical degrees of freedom. Strong coupling may occur with one choice and be absent with another. A striking example is the atomic nucleus: The two main basic models assume completely opposite approaches to describing the structure of the nucleus and the properties of nuclear matter. On the one hand, a good description of such collective properties of the nucleus as, for example, the binding energy is given by the droplet model, which assumes a strong interaction between the nucleons that make up the nucleus.
On the other hand, many properties of the spectra of atomic nuclei are best described within the framework of the shell model --- the simplest microscopic model of the nucleus, which assumes the motion of non-interacting nucleons in a self-consistent nuclear field. The standard constituent quark model of the nucleon is close in spirit to the shell model. Therefore, it cannot describe the mass of the nucleon, since this mass, in a first approximation, is simply the sum of the postulated masses of the constituent quarks.
In reality, the current quarks of the Standard Model make up about 2\% of the nucleon mass, the rest of the mass arising from the gluon field binding these quarks within the nucleon. Thus, when constructing models for the nucleon mass, approaches are needed that are initially aimed at description of very large binding energy. The polaron approach proposed in this paper is a new exploratory step in this direction.

The way we obtained the mass of the nucleon suggests a certain idea for constructing polaron models of the nucleon. In traditional quark models, all valence quarks inside the nucleon are independent degrees of freedom. In the polaron approach, apparently only one valence quark should be considered as an effective physical degree of freedom, with the other two valence quarks belonging to the "environment", alongside sea quarks and gluons creating a kind of continuous background that interacts with the "physical" quark neutralizing its color charge and changing its electrical charge. Thus, in this model of the nucleon, there is only one constituent quark strongly coupled to the quark-gluon background. By isospin symmetry, the "constituency" can be assigned to any of the three valence quarks. This dynamical picture of the nucleon sharply contrasts with the static picture of three feebly interacting constituent quarks given by the standard quark model.
It would be interesting to propose a model for the nucleon wave function within the polaron approach. This wave function should be based on a certain model of quark hadronization. Recall that
one of the most direct manifestations of the fractional electrical charge $Q_{u,d}$ of quarks was obtained in deep inelastic scattering:
In the limit $x\rightarrow1$, the ratio of neutron to proton structure functions $F_2(x)$ approaches~$\frac14$,
\begin{equation}
%\label{dis}
\lim_{x\rightarrow1}\frac{F^n_2(x)}{F^p_2(x)}\rightarrow\frac14=\frac{Q^2_d}{Q^2_u}.
\end{equation}
In the formal analysis of deep inelastic scattering, this translates into the statement that in the valence regime $x\rightarrow1$, where the struck parton carries all the longitudinal momentum of the proton,
that struck parton must be a $u$-quark~\cite{wil}. By isospin symmetry, the parton within a neutron in the same valence regime must be a $d$-quark.
Since the strong interactions are $T$-invariant, this implies also that if an energetic $u$-quark is injected into vacuum then it will be hadronized into a proton (of course, provided that hadronization ends with the formation of a nucleon). Correspondingly, an energetic $d$-quark in the QCD vacuum becomes a neutron. Note that a similar phenomenon happens with a free moving electron placed in ionic crystal --- it is effectively converted into a polaron with some ground state energy in a self-induced potential well.

According to the modern lattice QCD calculation of Ref.~\cite{Yang:2018nqn}, a rest-frame decomposition of the proton's mass $M_p$ into contributions from various components consists from
three approximately equal parts: (i) quark kinetic + potential energy; (ii) gluon kinetic + potential energy; (iii) trace anomaly + quark current-mass term (i.e., contributions from the quark scalar condensates). The first part (i) can be interpreted as the effective dynamical quark mass as a quantity related to the quark contribution. The last two parts (ii) and (iii) reflect contribution from the gluon sector (the contribution from the scalar condensates in (iii) emerges due to non-perturbative gluon interactions as well, see also footnote~3 below). Thus, one has a similar approximate proportion of $1:2$ for contributions to $M_p$ as in the polaron approach.

A heuristic way to understand the contribution (i) is the following~\cite{pm1}. If three massless quarks are confined to a spherical cavity of radius 1~fm, their
total kinetic energy will be of the order of $600$~MeV due to the uncertainty relation. Thus, the color-current interaction between quarks
and gluons contributes about $-300$~MeV to the mass, which is consistent with the magnitude of $N-\Delta$ splitting. Note that when interpreting (i) as the value of constituent quark
mass, we effectively ascribed this contribution to one of the valence quarks. It seems plausible that this replacement does not change noticeably the value of the nucleon mass, since this value gives a value for the internal energy that should be independent of any kinematic configuration of the internal degrees of freedom. It could be said that the proposed
polaron approach is partly based on this assumption.

The phenomenology of quark-gluon confinement has some counterparts in polaron physics. The following observations can be made.

At a very qualitative level, one can note a certain similarity between the properties of the LOP and the hadron string. In the case of the latter, lattice calculations in QCD confirm the following picture of the color confinement: The gluon field between two sources of color charge looks like a tube of the longitudinal chromo-electric field flux (see footnote~1), in which the chromo-magnetic component is absent --- the dual Meissner effect in the QCD vacuum. A possible polaron analogue of this phenomenon: The electric field $\vec{E}$ inside LOP is parallel to the wave vector $\vec{k}$ and has no magnetic component (which is associated with the transverse electric field).

It can be further noted that the LOP frequency $\omega_0$ in polaron theory is determined from the condition of zero permittivity,
$\varepsilon(\omega_0)=0$, which ensures the absence of the electric displacement vector $\vec{D}$ in the medium due to $\vec{D}=\varepsilon\vec{E}$. Such a transformation
of the electromagnetic field generated by the oscillations of the lattice ions, entirely into a longitudinal electric field inside LOP, can be interpreted as a polaron analogue of the gluon confinement. The polaron analogue of quark confinement consists in the ''confinement'' of a free electron inside the deformation of the ionic lattice caused by it --- as noted above, when fixing one of the valence quarks as an effective basic degree of freedom, it is useful to view the other valence quarks within the same nucleon as part of a "deformation" that creates a potential well, the ground state energy of which determines the mass of the nucleon.

Finally, we note another common property of the confinement of quarks in a hadron and an electron in a LOP --- the dominant contribution to both phenomena comes from the infrared region of interactions. In the second
case, the mechanism is as follows: An electron in an ionic lattice interacts most effectively with longitudinal optical oscillations whose wavelength is greater than the distance traveled
by the electron during the lattice oscillation period $2\pi/\omega_0$, since it is under this condition that the change in the crystal density necessary for the effect occurs and the formation of a polarization field of the medium
results in the electron being in a potential well all the time. If, due to some conditions (for example, under the action of a sufficiently strong external electric field),
the electron begins to travel a distance exceeding the phonon wavelength during the time $2\pi/\omega_0$, then the intensity of the electron-phonon interaction begins to rapidly decrease. This
can be interpreted as a polaron analogue of asymptotic freedom in QCD.

\section{Conclusions}

We have motivated a possible correspondence between the physics of longitudinal optical polarons arising in ionic crystals and low-energy strong interactions describing constituent quarks inside nucleons. Under the formulated rules of this correspondence, it leads to certain quantitative predictions that agree well with the hadron phenomenology. The most surprising result is the quantitative prediction of the nucleon mass with unexpected accuracy. Moreover, quantitative agreement is achieved using the improved result~\eqref{8} for the ground state energy of the polaron in the strong coupling limit.
The presented observations may imply that the quantum phenomena in the strong coupling regime discussed in this work likely possess a certain universality.

The traditional quark model provides, in the first approximation, a static picture of the nucleon (three weakly coupled quarks, each with a mass about $1/3$ of the mass of a nucleon), which is convenient for systematization of the nucleon states and description of the nucleon excitations. In this approach, the nucleon mass as a function of the constituent quark mass scales as $M_N\sim m^*_q$. The polaron model predicts the scaling $M_N^2\sim m^*_q$.
We argued that such a scaling is naturally expected in a relativistic approach based on the energy-momentum tensor in QCD.

Within our interpretation of the polaron approach, the nucleon as a composite quantum system is separated into one "physical" quark with a mass of about $1/3$ of the nucleon mass and "environment" containing two other valence quarks plus virtual gluons, pions and sea quarks. The "physical" quark is strongly coupled with this "environment" that gives about $2/3$ of the nucleon mass.
The approximate proportion of $1:2$ for contributions to the nucleon mass from the quark and gluon sector agrees with the modern lattice calculations~\cite{Yang:2018nqn}. It is interesting
to note that this proportion is preserved at high energies, where heavier quark flavors contribute~\cite{Hoferichter:2025ubp}.

Perhaps the main advantage of the presented approach is its simplicity --- no serious calculations are required to obtain results since the calculations are already performed in the polaron theory. We hope that this simplicity partly compensates for the speculative nature of some of the made assumptions.

Thus, a conceptually new model is proposed that links previously unrelated areas --- polaron theory and strong interactions. Despite its obvious simplifications, the model yields qualitatively and quantitatively reasonable results and may deserve further development.
For example, one can ask the following questions: What does the wave function of a nucleon look like in this approach? What new insight  could it give for the description of nucleon structure? What is the relativistic generalization of the used approach, taking into account the fermion spin? Are there any analogs in strong interactions for other types of polarons?
The construction of working models in these directions may be of interest as a new way of describing the phenomenology of non-perturbative strong interactions.

\section*{Appendix}

The gravitational mass of a stable hadron state $|h\rangle$ is defined from the trace
of QCD energy-momentum tensor $\Theta_{\mu\nu}$ in the rest frame (see, e.g.,~\cite{al,pm1,pm2,pm4,pm5,pm6})
%, below the sign convention is mostly positive),
\begin{equation}
\label{a1}
m_h=\frac{1}{2m_h}\langle h(p)|\Theta^\mu_\mu(0)|h(p)\rangle,
\end{equation}
where the denominator comes from the relativistic normalization of states, $\langle h(p)|h(p)\rangle=2p_0$.
%$\langle h(p)|h(p)\rangle=2p_0(2\pi)^3\delta^{(3)}(0)$.
The trace $\Theta^\mu_\mu$ is given by the scale anomaly in QCD plus the quark mass contribution,
\begin{equation}
\label{a2}
\Theta^\mu_\mu=\frac{\beta(g_s)}{2g_s}G_{\mu\nu}^2+\!\!\sum_{q=u,d,\dots}\!\!\!\!(1+\gamma_{m_{q}}) m_q\bar{q}q.
\end{equation}
Here $\beta$ denotes the QCD beta-function, $g_s$ is the strong coupling, $m_q$ is the mass of the quark $q$ and $\gamma_{m_q}$ represents the anomalous dimension of the quark mass operator.
The hadron mass has two renormalization scale independent contributions,
\begin{align}
\label{a2b}
m_h  =&  \frac{1}{2m_h} \langle h(p) | \frac{\beta(g_s)}{2g_s}G_{\mu\nu}^2 + \!\!\sum_{q = u, d, \dots}\!\!\!\! \gamma_{m_q} m_q \bar{q} q   | h(p) \rangle \nonumber \\
 &+  \frac{1}{2m_h}\langle h(p)|\!\!\sum_{q = u, d, \dots}\!\!\!\! m_q\bar{q} q  | h(p) \rangle.
% \label{a2c}
\end{align}
How to extract the hadron mass from the expression~\eqref{a2b} is a controversial question in the literature~\cite{pm1,pm2,pm4,pm5,pm6}.
For the purpose of this work, we should propose some outline of a formal scheme for deriving the nucleon mass from~\eqref{a2b}.
Our strategy will be the following: We find a set of assumptions which leads to GOR relation~\eqref{GOR} for the pion mass starting from definition~\eqref{a1} and after that apply these assumptions for a similar derivation of the nucleon mass from~\eqref{a1}.

Consider GOR relation~\eqref{GOR} in the $SU(2)$ isospin limit,
\begin{equation}
\label{GOR2}
m_\pi^2f_\pi^2=-2m_q\langle \bar{q}q\rangle.
\end{equation}
This relation is obtained from the SCSB in QCD with the pions playing the role of Goldstone bosons. There is no generally accepted derivation of~\eqref{GOR2} directly from~\eqref{a2b}.
We propose the following formal derivation. Take the $SU(2)$ isospin limit in~\eqref{a2b}, $q=u,d$. Since the pion mass disappears in the chiral limit, $m_q=0$, and is renormalization scale invariant, only the last term in~\eqref{a2b} should contribute,
\begin{equation}
\label{a2c}
m_\pi^2\simeq\frac12 \langle \pi(p)|\!\!\sum_{q = u, d}\!\! m_q\bar{q} q  | \pi(p) \rangle.
\end{equation}
This expression (modulo general factor) was implicitly contained already in the original GOR paper~\cite{Gell-Mann:1968hlm}.
There is one caveat~\cite{Zwicky:2023bzk,Hoferichter:2025ubp}: The Feynman--Hellmann theorem,
\begin{equation}
\frac{\partial E_\lambda}{\partial\lambda}=\langle\psi_\lambda|\frac{\partial \hat{H}_\lambda}{\partial\lambda}|\psi_\lambda\rangle
\end{equation}
where $\lambda$ is a parameter in the Hamiltonian, leads to the result that~\eqref{a2c} gives
exactly half the pion mass~\cite{Hoferichter:2025ubp}, 
\begin{equation}
\label{sigma}
\sigma_\pi\equiv\frac{1}{2m_\pi}\langle\pi|m_u\bar u u+m_d\bar d d|\pi\rangle = m_u\frac{\partial m_\pi}{\partial m_u}+m_d\frac{\partial m_\pi}{\partial m_d}=\frac{m_\pi}{2},
\end{equation}
where the relativistic normalization of states in~\eqref{a1} and GOR relation~\eqref{GOR} were used. This pion "$\sigma$-term" gives contribution to the pion mass arising from the matrix element of the quark scalar current in $\Theta^\mu_\mu$~\eqref{a2b}, while the rest comes from the first line of~\eqref{a2b}. The exact reason of this result is unclear, a possible scenario is the existence of infrared fixed point $g_s^*$ with vanishing beta-function, $\beta(g_s^*)=0$ (see discussions after relation~\eqref{10}), at which $\gamma_{m_{q}}=1$~\cite{Zwicky:2023bzk}.
%An alternative heuristic understanding of this scenario could be the following: The second line of~\eqref{a2b} represents an analogue of potential energy $U$, while the first line is %analogous to the kinetic energy $T$ (the quark masses there appear via quantum loops), and since $\langle T\rangle=\langle U\rangle$ for the harmonic oscillator (the ground state of quantum %fields is described as an infinite set of harmonic oscillators) one can take into account the "kinetic" contribution by simply doubling the "potential" part in~\eqref{a2c}.
However, these details will not bother us since below we will normalize the pion state by matching to GOR relation.

The averaging in~\eqref{GOR2} is taken over the QCD vacuum,
$\langle q\bar{q}\rangle\equiv\langle 0| q\bar{q}|0 \rangle$. The standard derivation of GOR relation~\eqref{GOR2} from~\eqref{a2c} (with doubled prefactor) is based on elimination of pions via a soft-pion theorem (a brief derivation is shown in~\cite{Zwicky:2023bzk}). However, it is not clear how to generalize this derivation to nucleons.
We need some alternative formal trick linking~\eqref{GOR2} and~\eqref{a2c} that can be extended to the nucleon case. Below we construct a heuristic link of $|\pi \rangle$ with $|0 \rangle$
in the sense that this link relates the averages,
\begin{equation}
\label{a2c2}
\langle \pi| \,\dots\,|\pi \rangle = a\langle 0| \,\dots\,|0 \rangle,
\end{equation}
with a known constant $a$.

Let us represent the projector $\hat{I}$ on a complete set of states in the form
\begin{equation}
\hat{I}=|0\rangle \langle0|+\dots,
\end{equation}
where $|0\rangle \langle0|$ is the projector on the QCD vacuum $|0\rangle$,
and insert $\hat{I}$ on both sides of $\bar{q} q$ in Eq.~\eqref{a2c}.
The pion mass is then given by
\begin{equation}
\label{a2d}
m_\pi^2\simeq \frac12 \langle \pi(p)|0 \rangle^2 \langle 0|\!\!\sum_{q = u, d}\!\! m_q\bar{q} q|0 \rangle+\dots
=\frac12 \langle \pi(p)|0 \rangle^2\langle\bar{q}q\rangle\sum_{q = u, d}\!\! m_q+\dots,
\end{equation}
where $u$ and $d$ quarks can be identified in the $SU(2)$ isospin limit. In other words, by assumption the operator of energy-momentum tensor couples to the state $|\pi\rangle$ in~\eqref{a2d} via virtual vacuum states plus some corrections. One can formally obtain GOR relation by assuming the hypothesis of vacuum dominance, $\hat{I}\approx|0\rangle \langle0|$, i.e., neglecting
corrections in~\eqref{a2d}.

Under the hadron wave function $\langle 0|h(p)\rangle$ we understand the sum of matrix elements over all possible operators which can create a state with the
quantum numbers of $|h\rangle$,
\begin{equation}
\label{a2e2}
\langle 0|h(p)\rangle=\langle 0|\sum_i C_h^i(p)\hat{O}^i_h|h(p)\rangle,
\end{equation}
where the coefficients $C_h^i(p)$ have such mass dimension that the sum is dimensionless. In the case of the pion, we again use the information that it is a pseudo-Goldstone boson, hence, the leading contribution
comes from the divergence of the axial current, $\hat{O}_\pi=\partial^\mu\! A_\mu$, that has the mass dimension equal to 4. The weak pion decay constant
(normalized to $f_\pi=92.2$~MeV) is defined by the matrix element
\begin{equation}
\label{a2h}
\langle 0|A_\mu(0)|\pi(p)\rangle=i\sqrt{2}f_\pi p_\mu,
\end{equation}
which results in
\begin{equation}
\label{a2i}
\langle 0|\partial^\mu\! A_\mu|\pi(p)\rangle=\sqrt{2}f_\pi m_\pi^2,
\end{equation}
where $m_\pi^2=p^\mu p_\mu$.  As follows from definition~\eqref{a2h}, the mass scale involved in the problem is $f_\pi |p_\mu|$, so to make $C_\pi(p) \partial^\mu\! A_\mu$
dimensionless it is natural to set
$C_\pi(p)=(f_\pi |p_\mu|)^{-2}=(f_\pi m_\pi)^{-2}$. We then arrive at
\begin{equation}
\label{a2g}
\langle 0|\pi(p)\rangle=\tilde{\imath}\frac{\sqrt{2}}{f_\pi},
\end{equation}
where the symbol $\tilde{\imath}$ is defined by
\begin{equation}
\tilde{\imath}\cdot\tilde{\imath}^*=-1.
\end{equation}
Substituting parametrization~\eqref{a2g} into~\eqref{a2d} leads to GOR relation~\eqref{GOR2} if we neglect, by using the vacuum dominance, the corrections.
Actually the major part of contributions from those corrections seems to be absorbed by the phenomenological value of the quark condensate in GOR relation.
The symbol $\tilde{\imath}$ is needed to ensure the correct overall sign in~\eqref{a2d} when passing from averaging over the bosonic state to averaging over
the vacuum state containing fermions.

A much shorter way of deducing~\eqref{a2g} can be done via the Feynman--Hellmann theorem.
Writing~\eqref{sigma} in the $SU(2)$ flavor limit as
\begin{equation}
\label{a2g2}
\langle\pi|m_q\bar{q}q |\pi\rangle=m_q\frac{\partial m_\pi^2}{\partial m_q},
\end{equation} 
and matching~\eqref{a2g2} with GOR relation~\eqref{GOR2}, we get the formal equality
\begin{equation}
\label{a2g3}
|\pi\rangle=\tilde{\imath}\frac{\sqrt{2}}{f_\pi}|0\rangle,
\end{equation}
which is equivalent to~\eqref{a2g} for the standard normalization of the vacuum state, $\langle 0|0\rangle=1$. The equality here must be understood in the sense of~\eqref{a2c2}.
The formal equality~\eqref{a2g3} can be interpreted as an expression of idea that the pions are collective excitations of the QCD vacuum due to their Goldstone nature. As a consequence, averaging over the vacuum state can be related to averaging over the pion states.

Now let us move on to the case of nucleons. Repeating the reasoning connecting~\eqref{a2c} with~\eqref{a2d}, we obtain the following relation for the mass of a nucleon in the rest frame,
\begin{equation}
\label{a3}
M_N^2=\frac12\langle 0|N\rangle^2\langle 0| \frac{\beta}{2g_s}G_{\mu\nu}^2+\!\!\sum_{q=u,d}\!\!(1+\gamma_{m_{q}}) m_q\bar{q}q|0\rangle.
\end{equation}
The one-loop QCD beta-function gives
$\frac{\beta}{2g_s}=-\frac{\beta_0}{8}\frac{\alpha_s}{\pi}$, where
$\beta_0=11-\frac23n_f$, $\alpha_s=\frac{g_s^2}{4\pi}$. Relation~\eqref{a3}
can be then rewritten as
\begin{equation}
\label{a4}
M_N^2\simeq\frac12\langle 0|N\rangle^2\langle\bar{q}q\rangle\left(- \frac{\beta_0}{8}\frac{\frac{\alpha_s}{\pi}\langle
G_{\mu\nu}^2\rangle}{\langle\bar{q}q\rangle} +\sum_{q=u,d} (1+\gamma_{m_{q}}) m_q \right).
\end{equation}
The quantity in brackets has the dimension of mass. In GOR relation~\eqref{GOR2}, the analogous quantity is equal to the sum of the quark masses.
If we represent this quantity in a similar way as the sum of effective masses of the quarks inside the nucleon,
\begin{equation}
\label{a5}
M_N^2\simeq\frac12\langle 0|N\rangle^2\langle\bar{q}q\rangle\sum_{q} m_q^*,
\end{equation}
a certain relation for the effective quark mass $m_q^*$ can be written. Neglecting contribution from the bare quark mass $m_q$ in~\eqref{a4} and assuming
that the nucleon consists of three effective valence quarks, we get
\begin{equation}
\label{a6}
m_q^*\simeq-\frac{\beta_0}{24}\frac{\frac{\alpha_s}{\pi}\langle
G_{\mu\nu}^2\rangle}{\langle\bar{q}q\rangle}.
\end{equation}
This relation emerged earlier in the model of Ref.~\cite{ahep}.
The effective gluon energy per each of three valence quarks in~\eqref{a6} can be estimated from the known phenomenology:
Substituting into~\eqref{a6} the standard value of gluon condensate from QCD sum rules,
$\frac{\alpha_s}{\pi}\langle G_{\mu\nu}^2\rangle=0.012(3)$~GeV$^4$~\cite{svz,svz2},
%and $\langle\bar{q}q\rangle=-(0.25\,\text{GeV})^3$, $n_f=2$,
we obtain the estimate (for $n_f=2$)
\begin{equation}
\label{a7}
m_q^*\simeq320\pm80\,\text{MeV}.
\end{equation}
The central value of $m_q^*$ in~\eqref{a7} is close to input~\eqref{cons} used in our work.
Note that approximately this value of mass should be assigned to the constituent $u$- and $d$-quarks~\cite{manohar} and which also commonly arises in lattice simulations~\cite{Aguilar:2012rz}.

Our interpretation of the polaron model for the nucleon formally corresponds to a different separation of the contributions in the brackets of relation~\eqref{a4},
\begin{equation}
\label{a8}
M_N^2\simeq\frac12\langle 0|N\rangle^2\langle\bar{q}q\rangle\left( -\frac{\beta_0}{12}\frac{\frac{\alpha_s}{\pi}\langle
G_{\mu\nu}^2\rangle}{\langle\bar{q}q\rangle} + m_q^* + \sum_{q=u,d} (1+\gamma_{m_{q}}) m_q \right),
\end{equation}
where $m_q^*$ is given by~\eqref{a6} and thus numerically coincides with the value of the standard constituent quark mass.
The expression in the brackets of~\eqref{a8} visualizes the main idea: The summation in the quark sector is over three valence quarks,
only one of which is constituent, plus the contribution of gluon energy in the approximate proportion of $1:2$.

We do not know how to deduce the value of the overall factor $\langle 0|N\rangle^2$. In the pion case, the value in~\eqref{a2g} is of the order of Compton wavelength
of the pion\footnote{Perhaps it would be more elucidating to use an alternative normalization for $f_\pi$ --- without $\sqrt{2}$ in~\eqref{a2h}, then $f_\pi=130.4$~MeV. Note that
$m_\pi\approx f_\pi+m_u+m_d$, i.e., the value of $f_\pi$ in this normalization reports the QCD contribution to the pion mass. Since this contribution is absolutely dominating, it defines
the inverse Compton wavelength $\lambda_\pi^{-1}$ of the pion. Thus, the definition~\eqref{a2h} tells us that the amplitude of the weak pion decay is inversely proportional to the effective linear size of the pion.}.
The nucleon is a composite system and its Compton wavelength is known to be determined by $m_\pi^{-1}$ rather than $M_N^{-1}$~\cite{Halzen}, so we expect $|\langle 0|N\rangle| = k/f_\pi$ with $k$ of the order of unity.
The overall factor $\langle 0|N\rangle^2$ stays in front of the interaction energy. In the perturbative domain, this energy is proportional to Casimir factor $C_F$ of $SU(3)$ color
gauge group, see~\eqref{cas}. If we assume that this property extends to the non-perturbative domain, then $\langle 0|N\rangle^2\sim C_F$. Since $C_F$ for baryons is half that for mesons, we would obtain from~\eqref{a2g} the factor $k=1$, i.e.,
\begin{equation}
\label{a9}
(?)\qquad\langle 0|N(p)\rangle=\frac{\tilde{\imath}}{f_\pi}.
\end{equation}
Substituting our guess~\eqref{a9} to~\eqref{a5} leads to the approximate relation~\eqref{11d2}.

Combining~\eqref{a9} with~\eqref{a2g} leads to the formal equality $|N\rangle=\frac{1}{\sqrt{2}}|\pi\rangle$, which can be understood as a formal expression of the heuristic idea that
the result of averaging over the nucleon states is proportional to averaging over the pion states, since
the nucleon at low energies looks like an object surrounded by a pion cloud, i.e. only these virtual pions are "visible" to an external probe.

In the nucleon case, the second line in~\eqref{a2b} gives the pion-nucleon sigma term, which is defined by~\cite{al}
\begin{equation}
\label{a10}
\sigma_{\pi N} =\frac{\overline{m}_{u,d}}{2 m_N} \langle N(p)| \bar{u}u + \bar{d}d  |N(p) \rangle.
\end{equation}
This term is renormalization scheme and scale independent and is usually interpreted as the light quark mass contribution to the nucleon mass.
According to this standard interpretation, the value of $\sigma_{\pi N}$ shows how much of the mass of ordinary matter is generated via the Brout-Englert-Higgs mechanism and how much is generated dynamically.
Leveraging our ansatz and~\eqref{a9} we obtain the estimate
\begin{equation}
\label{a11}
\sigma_{\pi N}\approx-\frac{m_q\langle \bar{q}q\rangle}{M_N f_\pi^2}.
\end{equation}
Substituting our input parameters we get a curious result: $\sigma_{\pi N}\approx 10\,\text{MeV}\approx m_u+m_d$. The existing phenomenological estimates lie in the interval
40--60~MeV~\cite{al}. We did not take into account the contribution from strange quark-antiquark pairs, this could be a cause of the discrepancy~\cite{al}.

\end{document}